\documentclass[sigconf]{acmart}
\renewcommand\footnotetextcopyrightpermission[1]{} 

\def\BibTeX{{\rm B\kern-.05em{\sc i\kern-.025em b}\kern-.08emT\kern-.1667em\lower.7ex\hbox{E}\kern-.125emX}}

\begin{document}

%
\title{A Virtual Obstacle Course within Diverse Sensory Environments}

\author{Zhu Wang}
\affiliation{%
 \institution{New York University}
 }

\author{Anat Lubetzky}
\affiliation{%
 \institution{New York University}
 }

\author{Charles Hendee}
\affiliation{%
 \institution{Tactonic Technologies}
 }

\author{Marta Gospodarek}
\affiliation{%
 \institution{New York University}
 }

\author{Ken Perlin}
\affiliation{%
 \institution{New York University}
 }

%
\renewcommand{\shortauthors}{Wang, et al.}

%
\begin{abstract}

We developed a novel assessment platform with untethered virtual reality, 3-dimensional sounds, and pressure sensing floor mat to help assess the walking balance and negotiation of obstacles given diverse sensory load and/or cognitive load. The platform provides an immersive 3D city-like scene with anticipated/unanticipated virtual obstacles. Participants negotiate the obstacles with perturbations of: auditory load by spatial audio, cognitive load by a memory task, and visual flow by generated by avatars movements at various amounts and speeds. A VR headset displays the scenes while providing real-time position and orientation of the participant's head. A pressure-sensing walkway senses foot pressure and visualizes it in a heatmap. The system helps to assess walking balance via pressure dynamics per foot, success rate of crossing obstacles, available response time as well as head kinematics in response to obstacles and multitasking. Based on the assessment, specific balance training and fall prevention program can be prescribed.


\end{abstract}

%
%
%



\maketitle
\section{Introduction}

Tripping over an obstacle and multitasking are two common causes of falls, and they are particularly challenging for aging adults\cite{bloem2001multiple, bergland1998falls}.
Often, aging adults fall while stepping onto a curb or over an obstacle either because they did not notice the obstacle or could not clear it safely. Aging adults may also have difficulty walking safely when engaged in a secondary task such as memorizing words\cite{lindenberger2000memorizing}.

Immersive VR environments can simulate real-life scenarios and help clinicians assess participants' balance conditions under controlled functional contexts\cite{meldrum2015effectiveness}.The functional contexts include different environments in which aging adults can experience sensory overload, such as a busy street. Difficulty in complex environments may prevent aging adults from participating in social activities. In a laboratory VR setup, participants can practice in a safe environment with no fear of actually experiencing a fall. Highly controllable and repeatable tasks in VR scenarios, together with the ability to create sensory load and cognitive load, may become a viable assessment of fall risk. Our system aims at assessing subjects' gait and ability to clear obstacles.

\begin{figure}[h]
  \centering
  \includegraphics[width=\linewidth]{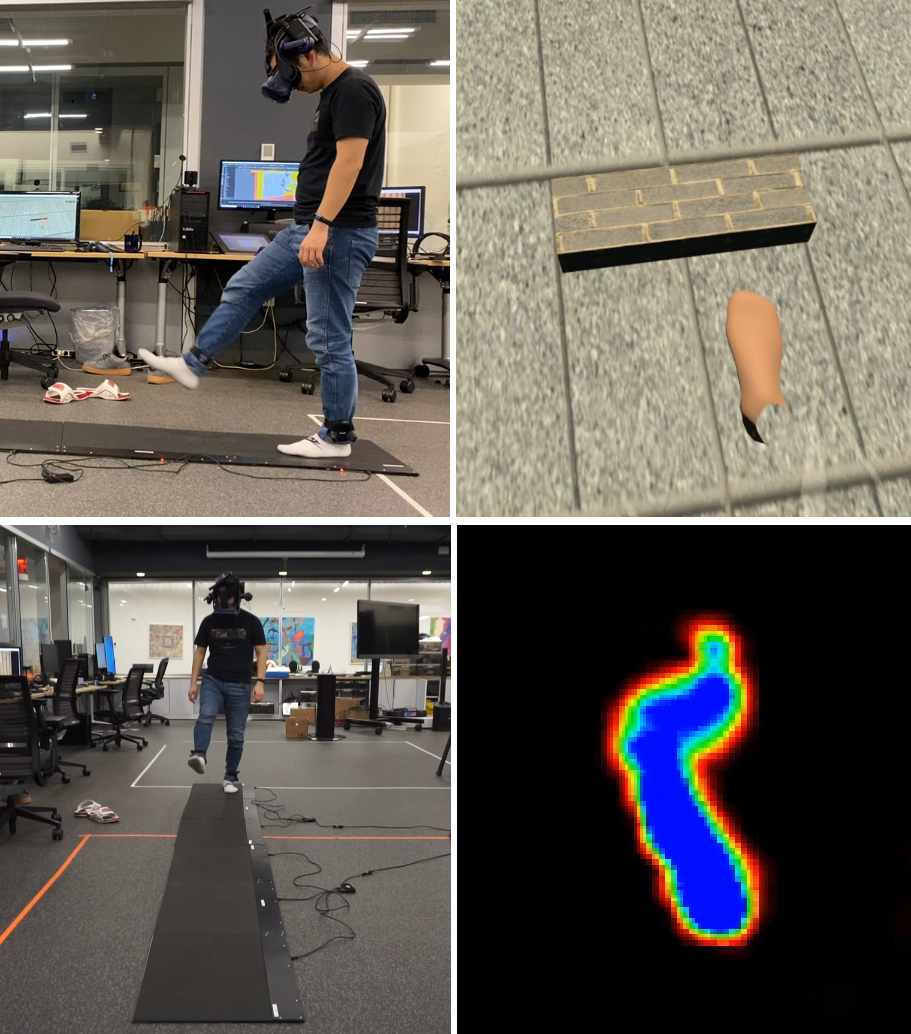}
  \caption{Front view and side view of the system setup, first persion view in VR, and visualization for pressure.}
  \label{fig:experiment}
\end{figure}

\section{System Design and Assessment}

We use the HTC Vive Pro with a wireless adapter to accomplish the untethered experience. With Vive trackers attached to the ankles, the system can track the participants' feet and detect if the feet collide with the virtual obstacles. Participants' feet are represented by human foot graphic models in VR. Thus, the participants will see that the foot models substitute their own feet. (see Figure \ref{fig:experiment})

The VR scene is a city-like scene which contains urban blocks with vehicles moving around, randomly generated buildings, virtual people on the street walking from one side of a sidewalk to the other side, and virtual obstacles generated on the sidewalk. Participants can cross virtual obstacles while walk along the virtual sidewalk during a 60-second session. We have tested the length of the walkway from 5 meters up to 15 meters due to the space requirement for a reasonable range of walking and the tracking capacity of the VR system. During the experiments, real-time position/orientation of the head and ankles, and pressure variances are obtained by the VR headset, Vive trackers, and the floor sensors respectively, so that we can further compare the head kinematics, foot clearance, obstacle negotiation strategy, success rate of crossing, and weight shifting under different intensity levels during participants' walking such as empty/busy street, and silent/bustling sound.

\subsection{Virtual Obstacles}

We assess subjects' balance when the subjects are instructed to cross the anticipated or unanticipated obstacles in the scene with auditory and visual stimuli. Clinicians can adjust the height of the obstacles among 25mm, 50mm, 75mm, 100mm, 125mm, 150mm, and 190mm. We chose 190mm to be the maximum height, because it is considered a standard stair based on the Stairbuilders and Manufacturers Association. The Vive trackers, which are mounted on the outside of the ankles, measure foot clearance in real-time. The system provides the participants with auditory feedback for success or failure of clearing obstacles.

In the anticipated obstacles experiments, all the obstacles are generated with selected height at the start of the scene. Participants can have enough time to plan the obstacle negotiation strategy. In the unanticipated obstacles experiments, all the obstacles are generated with selected height but at random time. Participants will not be able to predict how far or when an obstacle will appear in front of them. They have constrained time to respond to the suddenly appeared obstacles by adapting immediately to a new strategy and executing it to avoid tripping over and maintain a balanced gait pattern. In the unanticipated obstacle condition, we also quantify the available response time (ART)\cite{eyal2019successful}, which is the interval between the presence of obstacles and before the participant reaching it. The ART is correlated with the participant's walking speed and distance of presence in front of the participant. People with better walking balance ability usually have less ART, so we can use ART as a metric to quantify the walking balance ability.

\subsection{Multitasking}

In daily living, it is common that people do more than one task at the same time. We walk and talk, walk and try to remember something, walk and text, etc. Maintaining a stable gait while dual-tasking becomes more challenging for aging adults\cite{lindenberger2000memorizing}. Cognitive interference while walking was found to reduce gait speed and increase gait variability (i.e. stride to stride fluctuations) in older adults and individuals with neurological deficits. Moreover, a correlation between increased risk of falls and changes in attention-demands designated to walking was found in older adults' walking while talking \cite{beauchet2009stops, springer2006dual}.

Our system has a dual-task paradigm where participants are asked to walk and cross the obstacles while listening to a list of sentences containing numbers. The participants need to remember the numbers at the end of a 60-second walk. During each of the 60-second session, participants will hear 7 sentences randomly selected from 45 pre-recorded sentences. We can then quantify changes in gait parameters and obstacle-clearing under the dual-task condition.

\section{Pressure Sensing and Sound}

The Tactonic Technologies (TT) sensor system  is a force sensing matrix with sensing nodes distributed at regular 0.5 inch intervals across the grid. Each sensing node-distributed over the aforementioned 0.5 inch intervals-senses 12 bits of pressure from 50g of pressure up to 10kg of pressure. Each sensor system is 24 inches by 16.5 inches of active sensing area.
Multiple sensor systems can be arranged together to become a pressure sensing walkway.
The pressure-sensing walkway allow us to look at balance parameters for each cycle of gait between heel contact and toe off- center of force, force distribution between feet, step length/width, stride length, and foot angle.
When comparing to motion capture systems such as widely used passive optical MoCap system, pressure sensing technology has several unique advantages.
It quantifies the forces between the foot and ground during the stance phase by measuring the center of force and force distribution over time. The lower cost and easier setup further increase the potential outreach.


The implementation of sounds enhances the realism of the environments and the sense of immersion\cite{nordahl2014sound}, and allows to examine if participants' performance deteriorates with the presence of complex sounds. Soundscapes were designed using 3D audio technology, particularly dynamic binaural rendering and Ambisonics to enable natural sound changes depending on the position of the participant in the scene. They consist of two layers: ambience (background) sounds, providing general information about the environment the participant is in; and sound effects representing each moving object. Soundscapes were designed in 3D using HRTF rendering and/or Ambisonics technology to enable natural sound changes depending on the position of the participant in the scene. We have two levels of complexity/intensity. The levels vary in the number of sound sources in the scene, loudness, and spectral content.

\section{Contribution}
This novel system combines head-mounted display, 3D audio, head kinematics tracking, ankle tracking, and foot pressure sensing. The systems can be used to detect difficulty in obstacle-clearing when the obstacles are anticipated or unanticipated. We can then assess whether changes in performance are observed with: visual load, auditory load and/or cognitive load. This specific assessment can then guide fall prevention programs individualized to the participant's needs. The low cost of our platform will increase the outreach to communities that otherwise have limited access to such technology. A gap exists in the availability of ecologically valid but low-cost technology that can be widely available for assessment and treatment of participants with balance problems and fall risk. Our platform offers a unique solution to this problem.
\thispagestyle{empty}

\bibliographystyle{ACM-Reference-Format}
\bibliography{sample-base}

\end{document}